\documentstyle[12pt]{article}
\begin{document}

\title{Critical behavior for mixed site-bond\\ directed percolation}

\author{A.Yu. Tretyakov\\
School of Information Sciences, Tohoku University\\
Aramaki, Aoba-ku, Sendai, 980-77, JAPAN \vspace{3mm}\\
N. Inui\\
Himeji Institute of Technology\\
2167, Shosha, Himeji, 671-22, JAPAN
}

\maketitle

\setlength{\baselineskip}{6mm}
\begin{abstract}
\setlength{\baselineskip}{6mm}

We study mixed site-bond directed percolation on 2D and 3D lattices by
using time-dependent simulations. Our results are compared with
rigorous bounds recently obtained by Liggett and by Katori
and Tsukahara. The critical fractions $p_{site}^c$ and $p_{bond}^c$ of
sites and bonds are extremely well approximated by a relationship
reported earlier for isotropic percolation, $(\log p_{site}^c/\log
p_{site}^{c^*}+\log p_{bond}^c/\log p_{bond}^{c^*} = 1) $, where
$p_{site}^{c^*}$ and $p_{bond}^{c^*}$ are the critical fractions in
pure site and bond directed percolation.

\vspace{2cm}
classification numbers: 0250,0520,6460,6490

\end{abstract}

\clearpage

The mixed site-bond percolation is a natural extension of the
ordinary bond percolation and site percolation models. The model is
defined by opening sites of a lattice with probability $p_{site}$ and
bonds with probability $p_{bond}$, clusters are defined as combinations
of open sites connected by open bonds. The system is percolating in a
region of $p_{site}$ and $p_{bond}$ values in which an infinite
cluster exists.

The mixed site-bond percolation is a natural model for physical
phenomena in randomly restricted media (accounted for by closed sites)
with random iterations (accounted for by open and closed bonds).
Introduced by Reynolds et. al. \cite{rey}, it was used in treating
polymer gelation, capillary phenomena and in fracture theory (see
\cite{yan} and references in it) .

Current work is concerned with the directed version of the mixed
site-bond percolation model (Fig. 1), first introduced by Kinzel
\cite{kinz1,kinz}, in which bonds can be open only in a certain
direction, physically it can be accounted for by the existence of an
external field, restricting interactions allowed in the system (as in
the case of transport of particles in porous media, induced by an
external electrical field), or it could also be applied to
time-dependent phenomena, when movement in a chosen direction
corresponds to development in time (see \cite{kinz,durr,dpm}).

Both pure bond and pure site directed percolation are known to belong
to the same universality class known as DP (Directed Percolation)
universality class. The DP universality class is known to be very
robust; with a few exceptions all dynamical particle systems involving
extinction-survival transition belong to DP universality class (see
\cite{dp,dp1} and references in it). The mixed directed side-bond
percolation should be expected to belong to the DP universality,
although no numerical check to confirm it has ever been conducted.

Recently Katori and Tsukahara \cite{kat} proposed a lower boundary for
the phase transition line formed by $p_{site}$ and $p_{bond}$ values,
at which the percolation transition is taking place in the case of
mixed directed percolation in 1+1 dimensions. On the other hand, exact
upper and lower bounds for the phase transition line in 1+1 dimensions
have been recently proposed by Liggett \cite{lig}.

Based on a heuristic argument, Yanuka and Englman \cite{yan} proposed
an interpolation formula for the phase transition line for ordinary
(non-directed) mixed site-bond percolation in any dimension, allowing
to calculate the phase diagram for the mixed percolation from the
critical fractions $p_{site}^{c^*}$ and $p_{bond}^{c^*}$ for pure site
and pure bond percolation, corresponding to the end-points of the
diagram. Because of the random spread on data, the Monte Carlo results
for the intermediate region reported by the same authors (for a number
of lattices in 2D and 3D) don't allow to see any systematic difference
from the phase transition line given by the formula, so that the
precision of the critical line given by the formula is unclear.

The principal aim of the current work is the evaluation of the phase
transition line for the directed mixed site-bond percolation in 1+1 and
in 2+1 dimensions using Monte Carlo simulations. For 1+1 dimensions we
confirm the validity of upper and lower bounds suggested by Katori and
Tsukahara \cite{kat} and by Liggett \cite{lig}, and, for both 1+1 and
2+1 dimensions, show that an interpolation formula, similar to the one
suggested by Yanuka and Englman \cite{yan} for the ordinary mixed
site-bond percolation, is applicable with high precision, although we
are able to detect small systematic deviations. The time-dependent
simulation approach used in the current work allows, simultaneously
with the transition line, the evaluation of dynamical critical
exponents for the mixed directed percolation. Our results imply that,
not surprisingly, the mixed case belongs to the same universality class
to which both pure bond and pure site directed percolation belong.

To decide the critical line, we use a time-dependent simulation
\cite{lsl}. Mixed directed percolation is regarded as a growth
process, starting from a single site, the ``origin''. For 1+1
dimensions, space and time are defined as coordinates of sites on a
square lattice perpendicular and parallel to the (1,1) diagonal, and
growth clusters are formed by sites that can be reached from the
origin moving through open sites and bonds in the positive time
direction (see Fig. \ref{fig_2d}). Similarly, for 2+1 case we consider
growth on a simple cubic lattice with time axis directed along the
(1,1,1) diagonal. This kind of growth can be realized by updating
states of sites of a triangular lattice (see \cite{grass} for a
discussion of the technique). We calculated the survival probability
P(t) as the fraction of realizations containing sites connected to the
origin at time $t$ (realizations, surviving at time $t$); the average
number of sites connected to the origin at time t, n(t), averaged over
all realizations; and the average square radius $ R^{2}(t) $, averaged
over surviving realizations.  On the critical line, for big enough
$t$, it is expected that these quantities are governed by power laws

\begin{equation}
\begin{array}{ccc}
P(t) & \propto &t^{-\delta} \\
n(t) & \propto &t^{\eta} \\
R^{2}(t) & \propto &t^{z}
\end{array}
\end{equation}

The results of the time-dependent simulation are most conveniently
represented as local slopes data (see \cite{lsl} for a detailed
description), for example, for the number of sites connected to the
origin we plot $\eta(t)\equiv\ln[n(t)/n(t/m)]/\ln[m]$ against $1/t$,
taking m=7. For $p_{site}$ and $p_{bond}$ on the critical line the
true $\eta$ value is given by the intersection of the local slopes
curve with ordinata, while in off-critical simulation the local slopes
curves diverge for small $1/t$. Survival probability $P$ and the mean
square radius $R$ are treated in a similar way. Although the local
slopes curves for $R$ are quite insensitive to a change in $p_{site}$
and $p_{bond}$ values and thus less useful in determining the critical
line than in the cases of $n$ and $P$, we used the mean square radius
data to check the consistency of our simulation and to estimate the
$z$ values.

An example of local slopes curves for time dependent simulation is
given in Fig. \ref{fig_sl}. The data corresponds to pure site directed
percolation, and allows us to support the critical point value
provided by Onody and Neves \cite{onody} using series technique as
opposed to the result obtained by D.ben-Avraham et.al. \cite{avr}
using the transfer matrix technique. We only show the local slopes
curves for the mean number of sites connected to the origin. Here, and
in the rest of the current work, we take 1000 steps and we average
over $10^5$ configurations. The result strongly implies that the the
critical value obtained by Onody and Neves is correct, while the value
provided by D.ben-Avraham et.al. \cite{avr} seems to be off the
critical point by more than the error margin, given in \cite{avr}.

The overall results are given in Tab. \ref{tab_1} and in Tab.
\ref{tab_2}, all $p_{bond}^c$ values given have an uncertainty of
$\pm0.0002$. The uncertainty has been determined by making sure that
the local slopes curves for $n$ and $P$, taken at $p_{bond}$ value,
shifted from the one provided in Tab. \ref{tab_1} and Tab.\ref{tab_2}
by the value of the error margin, clearly diverge as $1/t$ approaches
zero.

Repeating the argument proposed by Yanuka and Englman \cite{yan} in
the case of mixed directed percolation, we have
\begin{equation}
\frac{\log p_{site}^c}{\log p_{site}^{c^*}}+\frac{\log p_{bond}^c}{\log
p_{bond}^{c^*}} = 1,
\label{eq_int}
\end{equation}
where $\log p_{site}^{c^*}$ and $\log p_{bond}^{c^*}$ are the critical
fractions of bonds and sites in pure bond and site directed
percolation.

We compare the critical line given by (\ref{eq_int}) with our Monte
Carlo results in Fig. \ref{fig_cmp_1} and Fig. \ref{fig_cmp_2}. The
formula works surprisingly well, although if the difference between
the interpolation formula prediction and the simulation results is
plotted separately, as is done in Fig. \ref{fig_cmpd}, one can see
that it can not be explained by the flaws of the Monte Carlo
simulation. Error bars, given in Fig.  \ref{fig_cmpd}, are calculated
as a combination of the uncertainty of the interpolation formula
prediction due to the uncertainty of the threshold values for pure DP
cases and of the error in $p_{bond}$ values, obtained in our
simulations. The critical values for pure DP cases were taken as
$p_{site}^{c^*}=0.705489\pm 0.000004$ \cite{onody} and
$p_{bond}^{c^*}=0.644701\pm 0.000001$ \cite{bondpc} in 1+1 dimensions,
and for 2+1 dimensions we used the values obtained by \cite{grass},
$p_{site}^{c^*}=0.43525\pm 0.00013$ and $p_{bond}^{c^*}=0.38216\pm
0.00006$.

Apart from the interpolation formula, in Fig. \ref{fig_cmp_1} we
compare the Monte Carlo results with exact boundaries for 1+1 mixed
directed percolation, proposed by Katori and Tsukahara and by Liggett.
Clearly, our results are in agreement with the bounds, presumed to be
exact.

Concerning the universality of mixed site-bond directed percolation,
both in 1+1 dimensions and in 2+1 dimensions, the dynamical exponents
$\delta$, $\eta $ and $t$, obtained in our time-dependent simulations,
were in agreement with the known DP values. For 1+1 dimensions, taking
the DP exponents values given in \cite{dp}, the error margins,
estimated by analyzing the spread of the local slopes data, were
given as: $\delta=0.162\pm 0.005$, $\eta=0.308\pm 0.005$ and
$z=1.263\pm 0.010$. For 2+1 dimensions, using the DP exponents values
from \cite{grass}, the error margins were given as $\delta=0.460\pm
0.020$, $\eta=0.214\pm 0.025$ and $z=1.134\pm 0.010$.  Thus, our
results indicate that mixed directed percolation in 1+1 dimensions and
in 2+1 dimensions belongs to the corresponding DP universality
classes.

Concluding the results of our work, we would like to note that the
interpolation formula suggested by Yanuka and Englman \cite{yan} for
the ordinary mixed site-bond percolation works surprisingly accurately
in the case of the directed mixed site-bond percolation as well.
Nonetheless, the accuracy of our Monte-Carlo results allows us to show
a systematic difference. Our results in 1+1 dimensions are in agreement
with exact bounds suggested by Katori and Tsukahara \cite{kat} and by
Liggett \cite{lig}. If the interpolation formula could be proven to be
an upper boundary for the critical line as our results suggest, it
would be a much closer bound than the Liggett's bound, the best
available so far.

\clearpage
\section*{Figure captions}
\begin{enumerate}

\item
Mixed directed site-bond percolation in 1+1 dimensions.
Percolation is allowed in directions indicated by the arrows. One
can clearly see a finite cluster, consisting of 9 sites, connected
to the origin.
\label{fig_2d}

\item
1+1 dimensions, local slopes for the mean number of sites connected
to the origin at different values of $p_{site}$, $p_{bond}=1$ for
all curves. Line shows the DP value of the exponent $\eta$, taken
from \cite{dp}.\\
{\bf 1-} $p_{site}=0.706522$ (critical value according to \cite{avr}),
{\bf 2-} $p_{site}=0.7057$,
{\bf 3-} $p_{site}=0.705489$ (critical value according to \cite{onody}),
{\bf 4-} $p_{site}=0.7053$,
{\bf 5-} $p_{site}=0.7051$.
\label{fig_sl}



\item
Critical line in 1+1 dimensions. Monte Carlo results are shown as
points.\\
{\bf 1-} The critical line given by Eq. \ref{eq_int}.\\
{\bf 2-} The lower bound by Katori and Tsukahara \cite{kat}, given as
a solution of $\alpha^3\beta^4-\alpha\beta^2+2\alpha\beta=1$ with
$\alpha=p_{site}^c$ and $\beta=p_{bond}^c$. \\
{\bf 3-} The upper and lower bounds by Liggett \cite{lig}, given by
$\alpha=\frac{2+\beta}{4\beta}$ and $\alpha=\frac{2}{\beta(4-\beta)}$,
respectively.
\label{fig_cmp_1}

\item
Critical line in 2+1 dimensions, for a simple cubic lattice.
\label{fig_cmp_2}

\item
The difference between the predictions of Eq. \ref{eq_int} and the
Monte Carlo results. Crosses indicate the uncertainty. \\
a)1+1 dimensions. \\
b)2+1 dimensions, simple cubic lattice. \\
\label{fig_cmpd}

\end{enumerate}

\clearpage
\section*{Table captions}
\begin{enumerate}

\item
Critical line in 1+1 dimensions, Monte Carlo results.\\
$^*$Values for ordinary directed percolation are taken from
\cite{bondpc} and \cite{onody}.
\label{tab_1}

\item
Critical line in 2+1 dimensions, for a simple cubic lattice.
Monte Carlo results. \\
$^*$Values for ordinary directed percolation are taken from
\cite{grass}.
\label{tab_2}

\end{enumerate}

\clearpage
\section*{Table \ref{tab_1}}
\begin{tabular}{ll}
$p_{bond}^c$&   $p_{site}^c$\\
$0.644701^*$  &   1\\
0.7       &   0.93585\\
0.75      &   0.88565\\
0.8       &   0.84135\\
0.85      &   0.80190\\
0.9       &   0.76645\\
0.95      &   0.73450\\
1         &   $0.705489^*$\\
\end{tabular}

\vspace{4cm}

\section*{Table \ref{tab_2}}
\begin{tabular}{ll}
$p_{bond}^c$&   $p_{site}^c$\\
$0.38216^*$ & 1 \\
0.45 &    0.86410\\
0.5  &   0.78725\\
0.6  &    0.67115\\
0.7  &    0.58770\\
0.8  &    0.52460\\
0.9  &    0.47510\\
1    &    $0.43525^*$
\end{tabular}

\end{document}